\documentclass[twocolumn,showpacs,preprintnumbers,amsmath,amssymb]{revtex4}

\usepackage{graphicx}
\usepackage{dcolumn}
\usepackage{bm}
\usepackage {amsmath}
\usepackage {amssymb}
\usepackage{subfig}

\begin{document}


\title{Hidden Tree Structure is a Key to the Emergence of Scaling in the World Wide Web}

\author{ZHENG Bojin(Ö£²¨¾¡)}
\email{zhengbojin@gmail.com} \affiliation{
 School of Software, Tsinghua University, Beijing 100084,China \\
 College of Computer Science, South-Central University for Nationalities, Wuhan 430074, China\\
}
\author{WANG Jianmin(Íõ½¨Ãñ)}%
\email{jimwang@tsinghua.edu.cn}
\affiliation{ School of Software, Tsinghua University, Beijing 100084,China \\}

\author{CHEN Guisheng(³Â¹ğÉú) }%
\affiliation{ Institute of Chinese Electronic Engineering, Beijing 100840,China \\}

\author{JIANG Jian(½­½¡) }%
\affiliation{Institute of Command and Technology of Equipment, Beijing 101416,China \\}

\author{SHEN Xianjun(ÉòÏÔ¾ı)}%
\affiliation{ Dept. of Computer Science, Central-China Normal University, Wuhan 430072,China \\}

\date{\today}

\begin{abstract}
Preferential attachment is the most popular explanation
for the emergence of scaling behavior in the World Wide Web, but this
explanation has been challenged by the global information hypothesis, the
existence of linear preference and the emergence of new big internet
companies in the real world. We notice that most websites have an obvious feature
that their pages are organized as a tree (namely hidden tree) and hence
propose a new model that introduces a hidden tree structure into the
Erd\H{o}s-R\'{e}nyi model by adding a new rule: when one node connects to another, it should also
connect to all nodes in the path between these two nodes in the hidden tree.
The experimental results show that the degree distribution of the generated
graphs would obey power law distributions and have variable high clustering
coefficients and variable small average lengths of shortest paths. The
proposed model provides an alternative explanation to the emergence of
scaling in the World Wide Web without the above-mentioned  difficulties, and also
explains the ``preferential attachment'' phenomenon.
\end{abstract}



\maketitle

For nearly 50 years, people have believed that most real networks are Poisson
random networks. This belief is based on the Erd\H{o}s-R\'{e}nyi model, which says that if
every node in a network is randomly connected to other nodes with a constant probability, the
distribution of degrees would obey a Poisson distribution. But Albert and
Barab\'{a}si\cite{1} found that some real-world networks, like webpages, have power law degree distributions, that imply mechanisms other than pure randomness. They proposed a model, which is called the BA (Barab\'{a}si-Albert) model, to explain the origins of power law degree distributions. In this model, every network would have initial seeds combined by a few nodes and the links
between them, and new nodes would connect to the existing nodes successively according to a linear probability that is  proportional to the degrees of existing nodes. This model is coined as preferential attachment mechanism, which is
similar to Yule Process\cite{28,30}, Price's model\cite{31}, and be a particular case of Simon
Model\cite{44,944}.

However, the BA model has been challenged for various reasons. First, in this model, every new node would know all
the information about the whole network\cite{46}. This hypothesis is called the global information hypothesis. As we know, this hypothesis is impossible in a large scale network because such huge quantities of data must be collected, stored, and processed. Second, every new node would make rational actions, not only attaching to other nodes
by preference, but also acting according to the linear rule, which may lead to a correlation between
degree and age of nodes\cite{49, 48, 50}. This hypothesis must also be unsatisfactory.
Lastly, when we apply the BA model to explain the
phenomena in the real world, we will encounter another problem. According to this model, webpages with higher connectivity become more and more important, i.e., the rich get richer, which means small internet companies would not be able to challenge existing large companies. Yet Google, Facebook, etc. are clear examples of websites that grew from very small to very large in spite of pre-existing large competing companies.

Scientists have proposed a couple of new complements for the BA model to respond to these challenges. For examples, Alexei
V\'{a}zquez proposed an ``adding + walking'' model to overcome the global information hypothesis\cite{46}.
Bianconi and Barab\'{a}si have proposed an extension to the BA model to explain why the age and the degree are not
correlated\cite{51,52}.
 Ravasz and Barab\'{a}si have proposed a model based on the hierarchical organization\cite{47}, this paper tries to put
 ``modularity, high degree of clustering and scale-free topology under a single roof''. But these models can
 not explain all these above-mentioned problems simultaneously.

Furthermore, preferential attachment is actually a variety of the Matthew effect\cite{96}. The Matthew
effect is actually a positive-feedback phenomenon. On the one hand, since it is a feedback, it probably could be accused of ``a vicious circle'', like the events happened to Darwinism(Some theologist, biologists and logicians think that there exist vicious circles in Darwinism, so the fossils are very important evidences to defense Darwinism. Preferential attachment mechanism can be regarded as the evolution theory in the World Wide Web, and this mechanism is similar to Darwinism with a positive feedback, so this mechanism may also be regarded as ``a vicious circle'' from pure logic view. Similar to Darwinism, it would need new evidences to be proved.); on the other hand, since it is a phenomenon, it may not be regarded as a basic fact. Therefore, we may need a more foundational fact to produce this effect and
then explain the power law distributions.

We notice that most websites organize their pages as a
tree, but this remarkable feature would not be easily reduced from the data of links. For example, the biggest Chinese website, www.sina.com.cn, organized its pages in many fields like the news, the sports and the finance etc., and the finance would include the stock, the money and the futures etc., and so on. Why do most websites organize their pages as trees?  It may owe to the fact that this kind of structure is easier to be understood by human visitors. In this circumstance, the links to identify the hierarchy play a role different from the other links in the World Wide Web, but all the links are not distinguished in the view of data mining since they have the same data format.  Because the hierarchy is not easily reduced from the data of links, we called it ``hidden tree'' following to ''Hidden Order'' coined by John Holland\cite{997}.

Because the hidden tree structures of websites are constructed for coinciding with the recognition of principles inherent to  humans, it would be reasonable that when one page is set to link to another, it probably would also be set to link to the pages in the shortest path from this page to the destination page in the hidden tree(In computer science, an explicit rule to design a good website says that a good website would not make visitors use the buttons in web browsers, that is, designers should set links to help visitors trace backwards.).

Based on the assumptions above, a new model is proposed to try to explain the emergence of scaling in the World Wide Web. In the proposed model, 1) all nodes are organized as a hidden tree, 2) the graph is generated by the following two rules: a) every node has a probability (called as $activity$) to connect to arbitrary and randomly selected nodes; b) when the source node connects to another destination node, it also connects to all the nodes in the path from the source node to the destination node in the hidden tree. This model is similar to the Erd\H{o}s-R\'{e}nyi model, since every node would connect to the others with an invariant probability, but one place is different, that is, this selected source node would also connect to the destination node and other nodes according to the hidden tree.

Based on the idea of ``hidden tree'', we assume that every
website has a hidden tree structure and all the hidden trees can be organized as only one hidden
tree by adding a virtual root node. So this model includes two parts: 1) the algorithm to construct the hidden tree; 2) the algorithm to generate the network according to constructed hidden tree. Since there exist many algorithm to generate a tree, so this paper does not focus on this problem; here this paper proposes an algorithm, which can be formally depicted as Fig.~\ref{fig:code}, to generate the network.

Assume that the hidden tree is an $n-tree$, where $n$ is the average children nodes of every node. When $n=2$ , it is a
binary tree. The number of nodes is denoted as $N$, and every node has a parameter $activity$. Moreover, we denote the adjacent
matrix as $AdjMatrix$ to represent the generating network, where the elements with value 1 represent
existing links. We use $SelectANodeRandomly()$ to represent the function to select a destination node randomly and we
use $GetPath()$ to represent the function to find all the nodes in the path from the source
node to the destination node in the hidden tree.

\begin{figure}[htbp]
\begin{flushleft}
0.Initialize AdjMatrix and generate the hidden tree;\\
1.for\ i=1:N \\
2.\ \ \ act = activity; \\
3.\ \ \ while act $>$ 0\\
4.\ \ \ \ \ \ rnd= random(0,1); \\
5.\ \ \ \ \ \ if rnd $<$ act\\
6.\ \ \ \ \ \ \ \ \ nodenum = SelectANodeRandomly(HiddenTree);\\
7.\ \ \ \ \ \ \ \ \ nodesInPath = GetPath(i, nodenum, HiddenTree);\\
8.\ \ \ \ \ \ \ \ \ AdjMatrix(i, nodenum)=1; \\
9.\ \ \ \ \ \ \ \ \ AdjMatrix(i, nodesInPath)=1; \\
a.\ \ \ \ \ \ end if\\
b.\ \ \ \ \ \ act = act -1;\\
c.\ \ \ end\ while\\
d.end\ for\\
\end{flushleft} \caption{\label{fig:code} The pseudocode of algorithm}
\end{figure}


For simplification, this model has set up some implicit assumptions. First, the hidden tree structure is
simplified as an $n-$tree. That is, every webpage would have the same number of children webpages.
Second, every webpage would have the same probability, i.e., $activity$ to link to the other pages.
Third, the hidden tree is static. Fourth, the rule to link to all the nodes in the path.
Under these assumptions, this paper only focuses on this problem: whether the hidden tree structure will produce a power law distribution with a ``preferential attachment'' phenomenon or not, such that to provide a candidate to the explanation to the emergence of
scaling in the World Wide Web.

This model is quite simple and has only three parameters. If this model can be used to explain the emergence of
scaling, the degree distribution of the generated networks would obey power law distribution, i.e. $P(k) \sim
 k ^{-\gamma}$, here $k$ is the degree and $\gamma$ is a constant.
The same as the BA model, we only discuss the in-degree distribution here. According to the experimental results,
we found that this model is very robust to all these three parameters.

To validate the effect of the number of nodes $N$, we choose that $n=2.0, activity =0.4$ and $N=1000, 2000,
5000, 10000, 20000$.

%

\begin{figure}
\begin{minipage}[t]{0.5\linewidth}
\centering
\includegraphics[width=4.72cm,height=3.6cm]{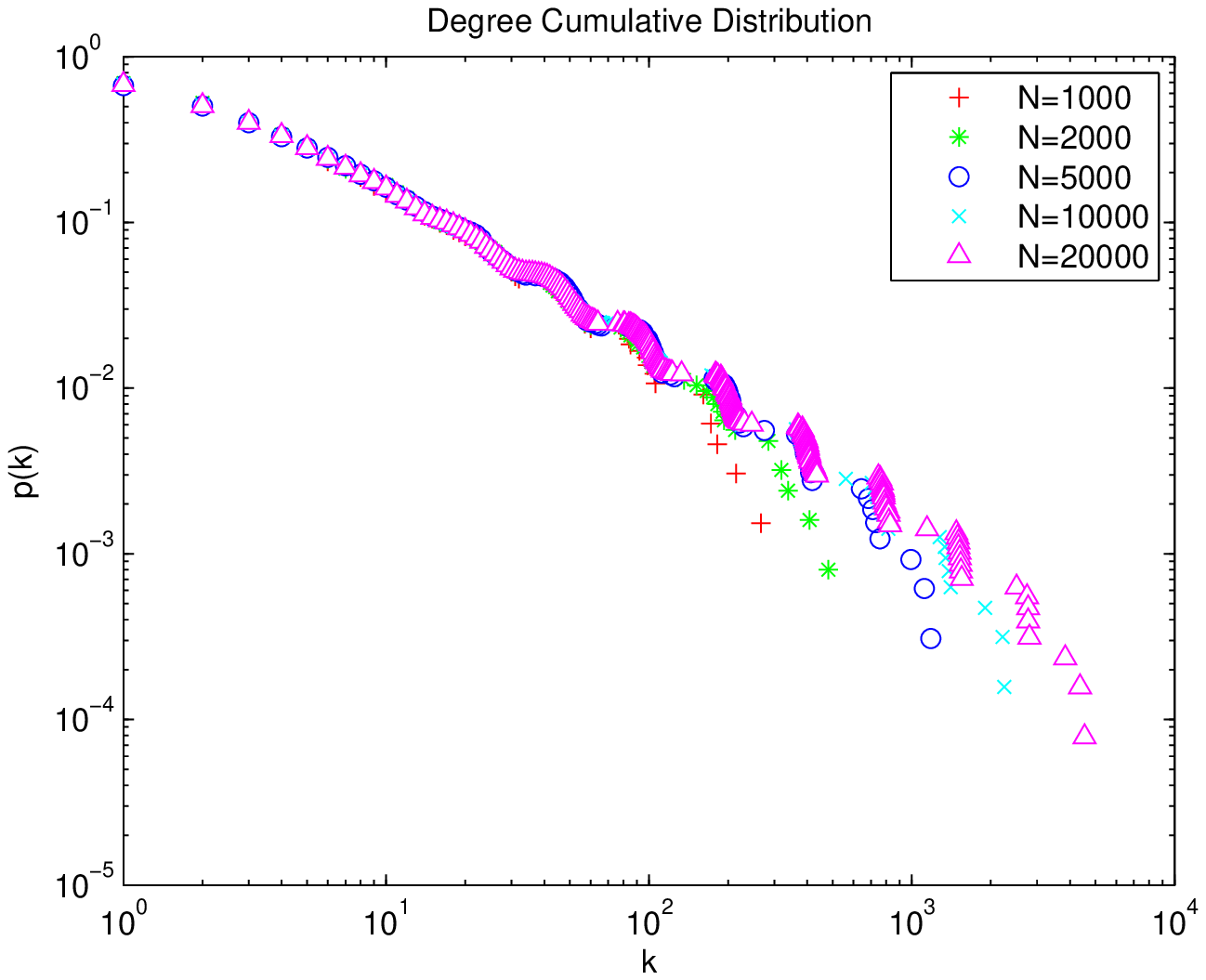}
\caption{Degree Cumulative Distribution When $N$ Varies}
\label{fig:NumNode}
\end{minipage}%
\begin{minipage}[t]{0.5\linewidth}
\centering
\includegraphics[width=4.72cm,height=3.6cm]{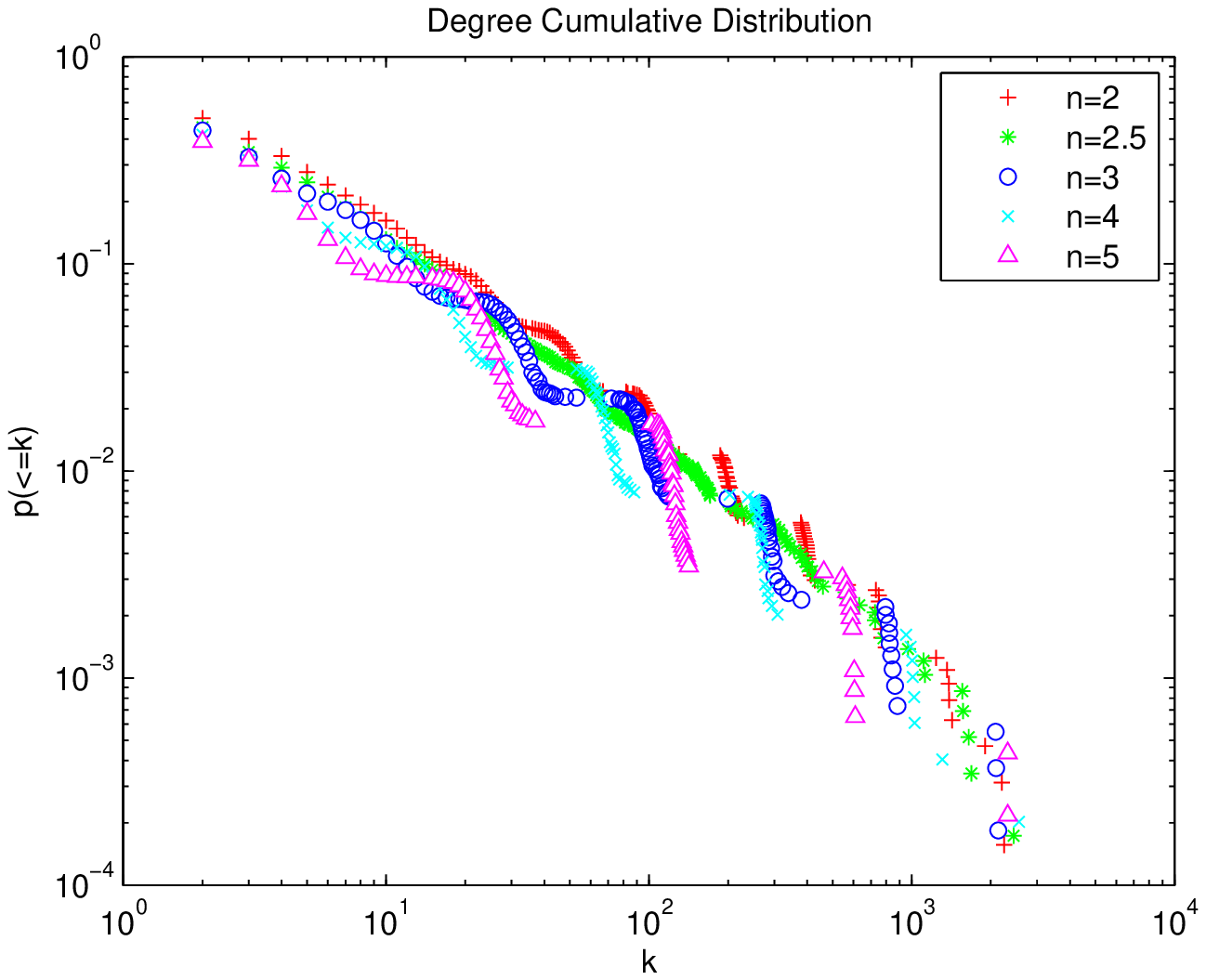}
\caption{Degree Cumulative Distribution When $n$ Varies}
\label{fig:children}
\end{minipage}
\end{figure}

From Fig.~\ref{fig:NumNode} we can see that when the number of the nodes varies, the exponent $\gamma $ does not
vary. At the tails of the curves, there is an exponential cutoff. According to the locations of the cutoffs,
when $N$ increases, the cutoff moves right. This phenomenon shows that the cutoff originates from
finite nodes. Moreover, from this figure, we can find that when $N$ increases, jumps emerge from the curves.
This effect may owe to discrete layers. Because of the randomness, the degrees of nodes in the same layer
would be different, resulting in the jumps in the curves.

To validate the effect the number of children $n$, we choose that $N=10000, activity = 0.4$ and $n = 1.5,
2.0, 2.5, 5.5, 7.5$.


From Fig.~\ref{fig:children} we can see that the curves are linear under loglog coordination, which make clear that in-degree
distributions satisfy power law distributions for different $n$.

To validate the effect of the parameter $activity$, we choose that $n=2.0, N = 10000$ and $activity = 0.08, 0.16,
0.32, 0.64, 1.28$(denoted as $A$).


\begin{figure}
\begin{minipage}[t]{0.5\linewidth}
\centering
\includegraphics[width=4.72cm,height=3.6cm]{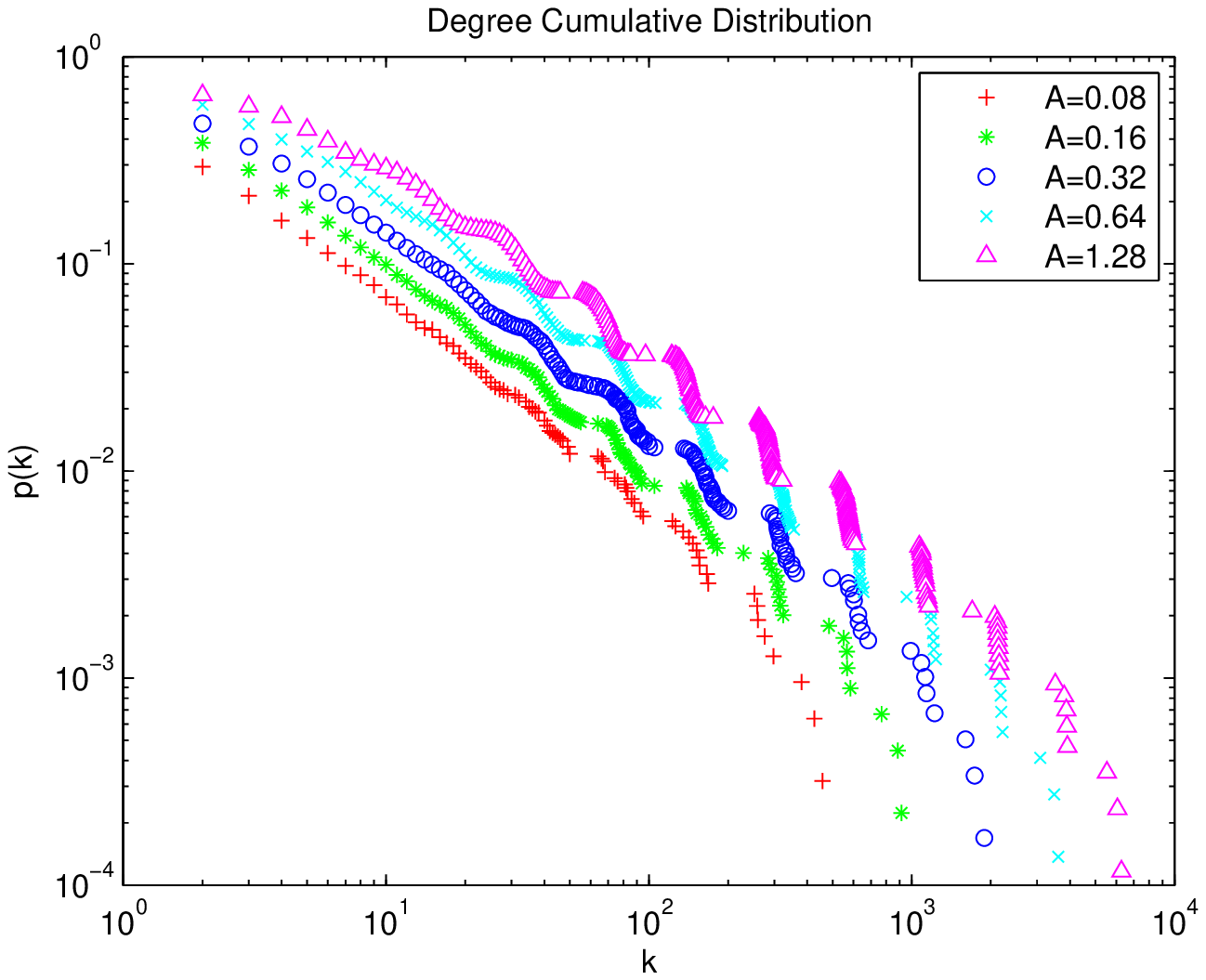}
\caption{Degree Cumulative Distribution When $activity$ Varies}
\label{fig:activity}
\end{minipage}%
\begin{minipage}[t]{0.5\linewidth}
\centering
\includegraphics[width=4.72cm,height=3.6cm]{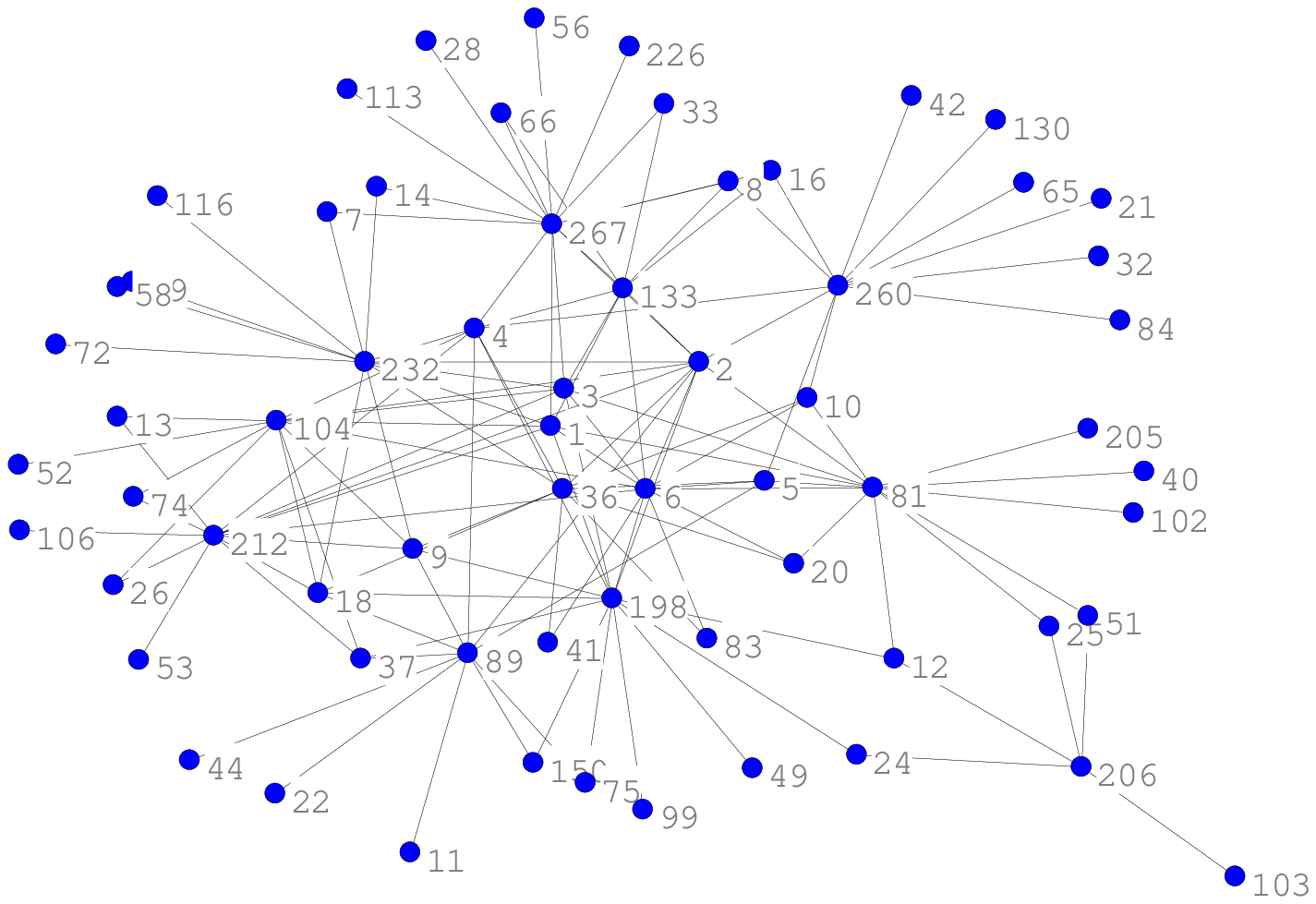}
\caption{Giant Component When $activity =0.04$}
\label{fig:CV1}
\end{minipage}
\end{figure}

From Fig.~\ref{fig:activity} we can see that the in-degree distributions also obey power law distributions for
different $activity$.

The experimental results above show that the in-degree distributions of the generated networks always obey power
law distributions regardless of different parameters.

If this model can explain the small-world effect, the generated network would have high clustering
coefficient and small average value of shortest paths. According to the next experimental results, we found
that the clustering coefficient is related to the parameter $activity$. When $activity$ increases slowly, the
pattern of the generated network initially is community-alike structure, secondly small world, finally super small
world. Of course, it would be a complete graph when $activity$ is large enough.

For visualizing the experimental results, we choose $N=300, n=2, activity = 0.04$, and $N=100, n=2, activity
=0.4$ and $2.0$.


\begin{figure}
\begin{minipage}[t]{0.5\linewidth}
\centering
\includegraphics[width=4.72cm,height=3.6cm]{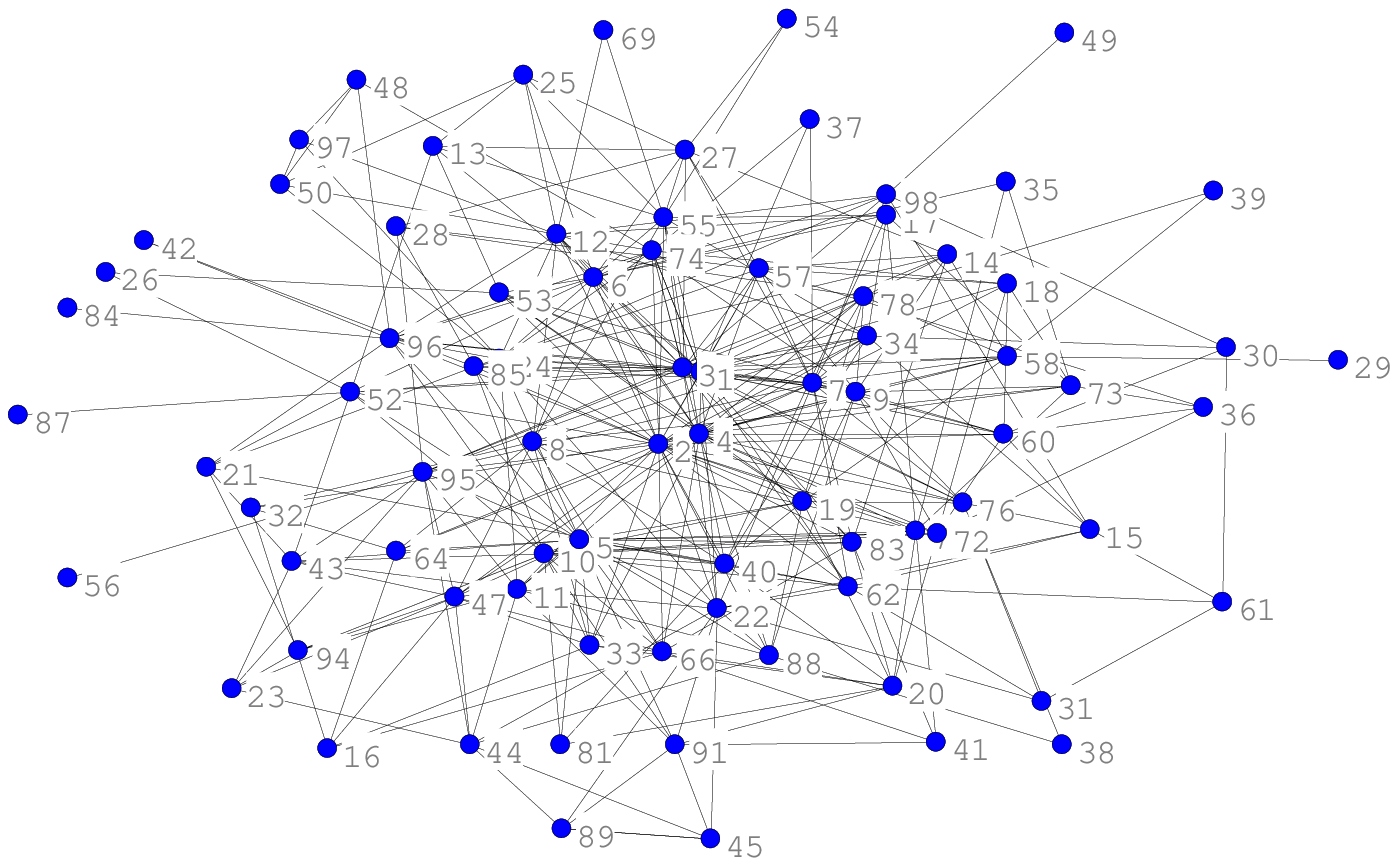}
\caption{Giant Component When $activity =0.4$}
\label{fig:CV2}
\end{minipage}%
\begin{minipage}[t]{0.5\linewidth}
\centering
\includegraphics[width=4.72cm,height=3.6cm]{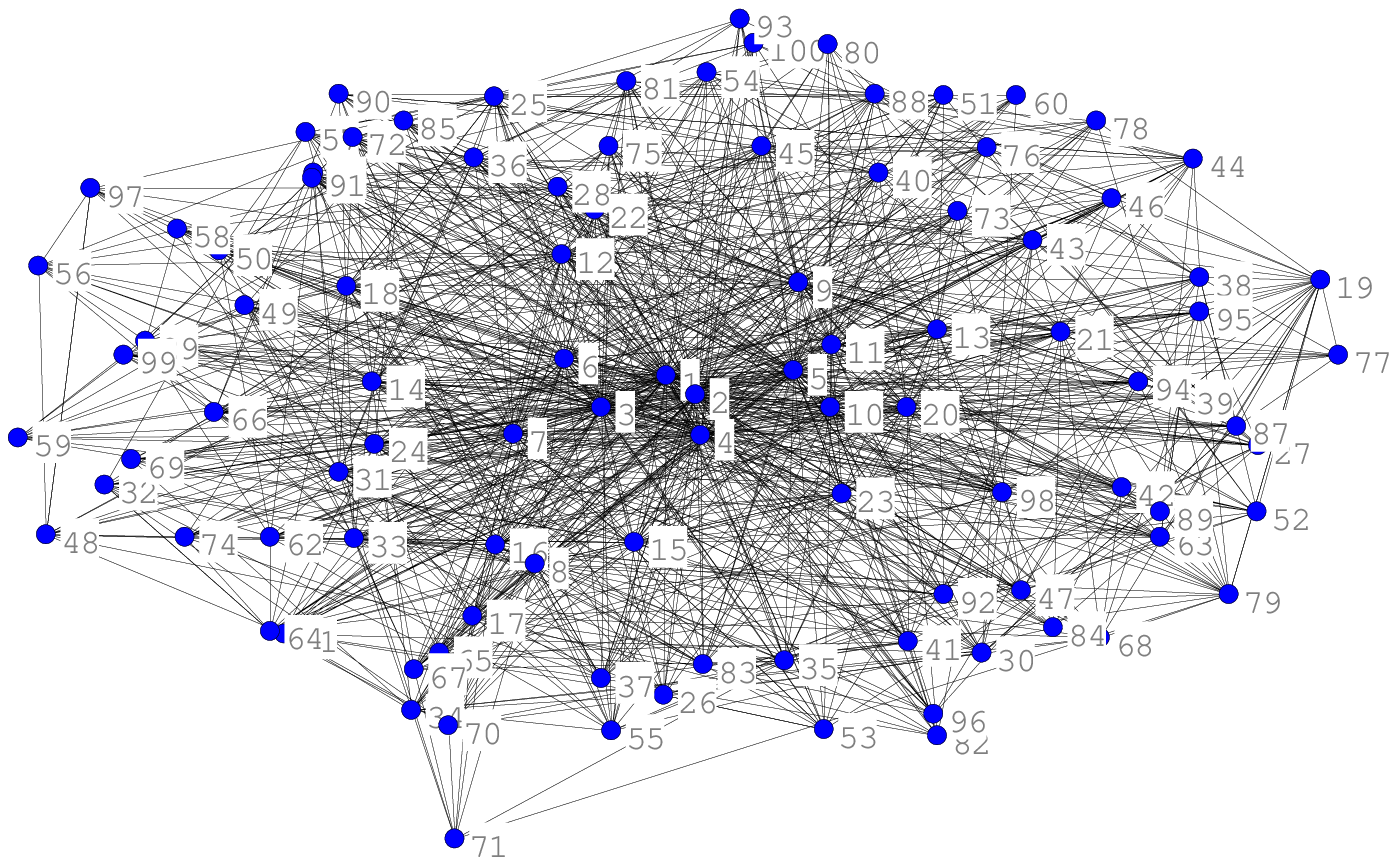}
\caption{Giant Component When $activity =2.0$}
\label{fig:CV3}
\end{minipage}
\end{figure}

%
%
%
%

From Fig.~\ref{fig:CV1} we can see that when $activity =0.04$, the giant component of the generated network looks
like to have community-alike structures. As to Fig.~\ref{fig:CV2}, the clustering coefficient increases
notably. Fig.~\ref{fig:CV3} is not a complete graph, but it has a larger clustering coefficient.
Moreover, the larger the average clustering coefficient are,  the smaller the average shortest paths are.
These results show that the small world effect does not conflict with the scale free property in this model.

In general, preferential attachment is widely accepted as the explanation of emergence of scaling in the World Wide Web.
However, challenges are difficult to be exiled by this mechanism itself. The proposed model employs an obvious feature in the World Wide Web,
hidden tree structure, and produces valuable results. First, the proposed model can also generate networks whose
degree distributions obey power law distributions. This conclusion implies that this model is a possible candidate to the
explanation of mechanism of power law distribution in the World Wide Web. Second, the proposed model has no the global information hypothesis, i.e., it is
unnecessary that every node would access all the nodes in the network. Third, the proposed model does not include
the obvious preference(In another view, the hidden tree can be regarded as a kind of preference). Instead of the preference, this model provides a reasonable
explanation that when one page connects to another, the creator of this link tends to assume that the visitors
would be interested in the generalized topic. Fourth, the proposed model integrates the small-world effect which
can be controlled by one parameter. Fifth, the proposed model can be used to explain the emergence of Google and
Facebook etc. Since no time-dependent relationship in the proposed model, the emergence of new companies means the insertion of subtrees, which may means ``niche''. Moreover, the proposed model has a useful feature that every node can take actions in a total parallelism, which means that the generated scale-free networks are ``stable'', independent of the time parameter of evolving process.

This proposed model does not refer to the ``preference'' directly, but it still includes the reasonable
ingredient of ``preferential attachment''. From the generated networks, for examples, Fig. \ref{fig:CV1}-Fig.\ref{fig:CV3}, we can find that the nodes with higher connectivity are closer to the center of networks, so we conclude that
the hidden tree structure model could be regarded as a possible foundational fact to some kinds of  ``Matthew
effects''.

We can also extend this model. For example, if
we set all the leaf nodes in the hidden tree active, that is, they can link to other nodes; and set the others inactive, that is,
they only accept the links, then this model could be easily reduced to a classic kind of combinations of
exponentials\cite{27,92,93,94}, and here the hidden tree structure becomes the perfect carrier of
exponentials. In general, regardless of these mechanisms, the hidden tree structure may be a key to the emergence of
scaling in the World Wide Web.

Though this model is proposed to explain the emergence of scaling in the World Wide Web, it may be also used to
explain food webs, wealth condensation etc. These phenomena include hidden trees too, and the metaphors in these fields would be very interesting. For instance, if hidden tree structure can produce power-law wealth distribution, according to the inverse theorem,  any actions on eliminating the uneven wealth distribution would be impossible, since it should destroy the hidden tree structure, which may mean social or economical structure, therefore lead to a serious economic degeneration. These results would be discussed in the future work.

\begin{acknowledgments}
The authors gratefully thank Prof. Deyi Li, Dr. Chunlai Zhou, Dr. Oskar Burger, Prof. JinHu L\"{u}, Mr. Haisu Zhang and Mr. Shuqing
Zhang for valuable discussions and helps. This work has been supported by the National Grand Fundamental
Research 973 Program of China under Grant No. 2007CB310804 and National Science Foundation of China under Grant No. 60803095 and SCUEC
Foundation under Grant No. YZZ06025.
\end{acknowledgments}

\end{document}